\documentclass[pra,twocolumn,showpacs,superscriptaddress,byrevtex]{revtex4}
\usepackage{amsfonts}
\usepackage{amsmath}
\usepackage{amssymb}
\usepackage[dvips]{graphicx}

\begin{document}

\title{Experimental scheme for unambiguous discrimination of \\linearly independent symmetric states}
\author{O. Jim\'enez}
\affiliation{Center for Quantum Optics and Quantum Information,
Departamento de F\'{\i}sica,  Universidad de Concepci\'{o}n,
Casilla 160-C, Concepci\'{o}n, Chile.}
\author{X. S\'anchez-Lozano}
\affiliation{Center for Quantum Optics and Quantum Information,
Departamento de F\'{\i}sica,  Universidad de Concepci\'{o}n,
Casilla 160-C, Concepci\'{o}n, Chile.}
\affiliation{Instituto de F\'isica Universidad de Guanajuato, P.O.
Box E-143, 37150, Le\'on Gto., M\'exico.}
\author{A. Delgado}
\affiliation{Center for Quantum Optics and Quantum Information,
Departamento de F\'{\i}sica,  Universidad de Concepci\'{o}n,
Casilla 160-C, Concepci\'{o}n, Chile.}
\author{C. Saavedra}
\affiliation{Center for Quantum Optics and Quantum Information,
Departamento de F\'{\i}sica,  Universidad de Concepci\'{o}n,
Casilla 160-C, Concepci\'{o}n, Chile.}
\email{carlos.saavedra@udec.cl}
\date{\today}

\begin{abstract}
We propose an optimal discrimination scheme for a case of four
linearly independent nonorthogonal symmetric quantum states, based
on linear optics only. The probability of discrimination is in
agreement with the optimal probability for unambiguous
discrimination among N symmetric states [Phys. Lett. A
\textbf{250}, 223 (1998)]. The experimental setup can be extended
for the case of discrimination among $2^M$ nonorthogonal symmetric
quantum states.
\end{abstract}


\maketitle

\section{INTRODUCTION}
\label{introduction}

A central problem in several quantum communication protocols, such
as quantum cryptography \cite{Bennett91,Ekert92,Bennett92},
quantum teleportation \cite{Bennett93} and entanglement
concentration \cite{Chefles98}, is the discrimination among
nonorthogonal quantum states, which can not be conclusively
discriminated with von Neumann's measurements alone. For
unambiguous discrimination among non-orthogonal quantum states it
is necessary to use generalized quantum measurements. The
discrimination process is error free in the case of a conclusive
measurement, where the probability of obtaining an inconclusive
result is non zero. In pioneer studies of Ivanovic-Dieks-Peres
(IDP) \cite{Ivanovic-Dieks-Peres}, for finding the optimal
probability of conclusive discrimination between two
non-orthogonal quantum states, with equal \textit{a priori}
probability, they found that:
\begin{equation}
P_{IDP}=1 - \vert\langle \Psi_+|\Psi_-\rangle\vert
\end{equation}
is the probability for error-free state discrimination, where
$|\Psi_+\rangle$ and $|\Psi_-\rangle$ are the states being
discriminated. Jaeger and Shimony have generalized this result by
considering the case of states with different \textit{a priori}
probabilities \cite{Jaeger}. Experiments for discriminating
between non-orthogonal polarization states at IDP limit were
accomplished utilizing linear optics only \cite{diez,once}. In the
same way, the experimental setup of unambiguous discrimination
among three non-orthogonal quantum states was carried out by
Mohseni \textit{et. al} \cite{Mohseni}, with a success rate of $55
\%$.

If we have a set of $N$ non-orthogonal quantum states denoted as
$\{|\Psi_k\rangle \}$, with $k=0,...,N-1$ lying in $N$-dimensional
Hilbert space $\mathcal{H}$, there exists no a general strategy
for unambiguous discrimination. If these states are linearly
independent, it is possible to conclusively discriminate among
them with a certain success probability. For this purpose, we need
to extend the $N$-dimensional space at most to a dimension $2N-1$.
This can be done by entangling the quantum system to a
two-dimensional ancillary system (ancilla) \cite{Roa}. After
coupling the ancilla to the quantum system, usually under
conditional evolution, a measurement over the ancilla projects the
quantum system onto a state which depends on the result of
ancilla's measurement. As we are dealing with a two dimensional
ancilla, one of the results will allow conclusive discrimination
of the original quantum state, and the other one gives an
inconclusive measurement. If operators $A_I$ and $A_k$ describe
the action on the quantum system in the cases of inconclusive and
conclusive results, respectively, they must satisfy the relation
\begin{eqnarray}
\sum_{k=0}^{N-1}A^{\dag}_kA_k + A^{\dag}_IA_I= \textbf{1}.
\end{eqnarray}
In this article, we study the problem of discriminating
non-orthogonal quantum states lying in a $2^M$ dimensional Hilbert
space. For sake of simplicity, we describe an experimental setup
in the case of dimension $4$ ($M=2$), which can be directly
generalized to larger dimensional cases. The setup considers the
generation process, propagation and discrimination of quantum
states. We restrict ourselves to the case of non-orthogonal
linearly independent states $\{|\Psi_k\rangle\}$ which are
symmetric, defined by:
\begin{equation}
|\Psi_l\rangle=Z^{l}|\Psi_0\rangle,
\end{equation}
where $|\Psi_0\rangle=\sum^{N-1}_{k=0}c_k|k\rangle$ is a
normalized state, i.e., $\sum_{k=0}^{N-1}\vert c_k\vert^2=1$. The
action of the $Z$ operator on this state is such that
$Z|k\rangle=exp(\frac{2\pi ik }{N})|k\rangle$ and $Z^N=I$. In ref.
\cite{CheflesBarnett} the action of the conditional unitary
evolution of a two-dimensional ancilla with the quantum system is
written as :
\begin{equation}
U|\Psi_l\rangle\otimes|0\rangle_a=\sqrt{p_l}|u_l\rangle|0\rangle_a+\sqrt{1-p_l}|\phi_l\rangle|1\rangle_a.
\nonumber
\end{equation}
where the $|0\rangle_a$ state is a known initial state of the
ancillary system, the states $\{|u_l\rangle\}$ and
$\{|\phi_l\rangle\}$ are orthogonal states and linearly dependent
states, respectively, of the quantum system. In the case of
measuring an ancilla in $|0\rangle_a$ state, the $|\Psi_l\rangle$
state is projected onto $|u_l\rangle$ state, with success
probability $p_l$, which allows a conclusive discrimination with a
von Neummann measurement in the basis $\{|u_k\rangle\}$, since
these states are orthogonal. In the case of the outcome
$|1\rangle_a$ for the ancilla, the state of the system is
projected onto linearly dependent states $\{|\phi_l\rangle\}$,
which can not be unambiguously discriminated. In this process the
optimal conclusive probability to discriminate between a set of
$N$ non-orthogonal symmetric states is $P_{opt} = N * min|c_k|^2$
\cite{CheflesBarnett}, where $c_k$ is the minimum coefficient,
i.e., $|c_k| \leq |c_l|$ of state $|\Psi_0\rangle$ for
$l=0,1,...,N-1$.

This article has been organized as follows: In Sec. II we
determine the conditional unitary transformation necessary for
discrimination in the case of four non-orthogonal symmetric
states. In Sec. III we describe an experimental setup for
generating, propagating and discriminating among the four
non-orthogonal states. This setup is based on down converted
photons generated in a spontaneous down converted (SPDC) process .
Finally, in Sec. IV we summarize our results and describe the
application of them to several quantum communications protocols.

\section{System-ancilla conditional evolution}
\label{trans}

Here, we consider the case of four non-orthogonal linearly
independent symmetric states, which are denoted by
$\{|\Psi_0\rangle, |\Psi_1\rangle, |\Psi_2\rangle,
|\Psi_3\rangle\}$. These states are generated by applying the
unitary transformation $Z^l$ onto the $|\Psi_0\rangle$ state, such
that $|\Psi_l\rangle=Z^l|\Psi_0\rangle$, with $l=0,1,2,3$. The
$|\Psi_0\rangle$ state is defined by:
\begin{equation}
|\Psi_0\rangle=\sum\limits_{k=0}^{3}c_k|k\rangle,
\label{eq:symstates}
\end{equation}
where the $c_k$ coefficients obey the normalization condition and
we will assume them to be reals. In general, these coefficients
can be written as $c_0 = cos\theta_1$, $c_1 = cos\theta_2
sin\theta_1$, $c_2 = cos\theta_3sin\theta_2sin\theta_1$ and
$c_3=sin\theta_3sin\theta_2sin\theta_1$. The convenience of this
notation becomes clear later on, when we discuss the physical
implementation of the discrimination protocol. For building up the
conditional unitary evolution, we will make use of the general
approach proposed by He and Bergou \cite{He}, which allows to find
a transformation that projects the $|\Psi_l\rangle$ states onto a
set of orthogonal states $\{|u_l\rangle\}$ and onto another set of
linearly dependent states, $\{|\phi_l\rangle\}$. Firstly, we must
get the diagonal form of $A^{\dag}_IA_I$ operators; this can be
done when there exits a unitary operator $U_o$ acting on the
initial Hilbert space which gives:
\begin{eqnarray}
U_oA^{\dag}_IA_IU^{\dag}_o=\sum_{i=0}^{D-1}\lambda_i|\alpha_i\rangle
\langle \alpha_i|,
\end{eqnarray}
where $|\alpha_i\rangle$ is an eigenvector of the $A^{\dag}_I A_I$
operator with eigenvalue $\lambda_i$. Since the $A^\dag_IA_I$
operator is positive, its eigenvalues are defined between zero and
one, and therefore we can define hermitian operators
\begin{eqnarray}
A^{\dag}_I=A_I=U^{\dag}_o\sum_{i=0}^{D-1}\sqrt {\lambda_i}|\alpha_i\rangle \langle \alpha_i|U_o,\\
A^{\dag}_s=A_s=U^{\dag}_o\sum_{i=0}^{D-1}\sqrt{1-\lambda_i}|\alpha_i\rangle
\langle \alpha_i|U_o.
\end{eqnarray}
The unitary transformation, in the enlarged space ancilla-system,
takes the following form:
\begin{eqnarray}
U=\begin{pmatrix}
 A_s&-A_I  \\
 A_I&A_s
\end{pmatrix}. \label{U}
\end{eqnarray}
where $A_s^{\dag}A_s=\sum_{k=0}^{N-1}A^{\dag}_kA_k$ is the
operator corresponding a conclusive result. The $U$ operator is
not unique, there are three other similar forms \cite{He}. We have
assumed a qubit ancilla, with basis $\{|0\rangle_a, |1\rangle_a\}$
and initially prepared in the state $|0\rangle_a$. After the
conditional evolution of the composite ancilla-system, the
measurement on the ancilla giving the state $|0\rangle_a$
determines the action of the $A^{\dag}_kA_k$ operator on the
original quantum system, so that the discrimination process is
conclusive. In the other case, the measurement on the ancilla is
$|1\rangle_a$, the POVM element $A^{\dag}_IA_I$ had acted on the
quantum system and hence the discrimination process fails. An
explicit form for the $A_k$ operator was found by Chefles
\cite{Chefles98},
\begin{eqnarray}
A_k=\frac{\sqrt{p_k}}{\langle
\Psi_k^{\perp}|\Psi_k\rangle}|u_k\rangle\langle \Psi_k^{\perp}|,
\end{eqnarray}
where the $|u_k\rangle$ states form an orthonormal basis for
$\mathcal H$; $|\Psi^\perp_k\rangle$ are the reciprocal states;
and $p_k$ is the probability to get the $k$-th outcome.  This
operator is consistent with
\begin{eqnarray}
A_k|\psi_k\rangle=\sqrt{p_k}|u_k\rangle.
\end{eqnarray}
The reciprocal states $|\Psi^{\perp}_k\rangle$ are defined by
\begin{eqnarray}
|\Psi^{\perp}_k\rangle=\frac{1}{\sqrt
q}\sum_{r=0}^{N-1}\frac{1}{c_r^*}e^{\frac{2\pi i}{N}kr}|r\rangle,
\end{eqnarray}
where $q=\sum_j\vert c_j\vert^{-2}$ \cite{CheflesBarnett}. These
states are also linearly independent and symmetric with respect to
the $Z$ transformation. Then operators $A_s$ and $A_I$ in case of
discriminating $\{|\Psi_0\rangle, |\Psi_1\rangle, |\Psi_2\rangle,
|\Psi_3\rangle\}$ states are:
\begin{eqnarray}
A_s &=& sin\theta_3sin\theta_2tg\theta_1 |0\rangle \langle 0 |  +
sin\theta_3tg\theta_2 |1\rangle \langle 1 | \\ &+& tg\theta_3
|2\rangle \langle 2 | + |3\rangle \langle 3 | \nonumber
\end{eqnarray}
and
\begin{eqnarray}
A_I &=& \sqrt{1-sin^2\theta_3sin^2\theta_2tg^2\theta_1} |0\rangle
\langle 0 |   \\ &+& \sqrt{1-sin^2\theta_3tg^2\theta_2} |1\rangle
\langle 1 | + \sqrt{1-tg^2\theta_3}  |2\rangle \langle 2 |.
\nonumber
\end{eqnarray}
Here, we have assumed that all \textit{a priori} probabilities
$\eta_k$ to be equal, with a value $\frac{1}{N}$ and the
discrimination probabilities to be $p_k = p_D$
\cite{CheflesBarnett}.

After applying the conditional evolution on the compound
ancilla-system, we get
\begin{equation}
U|\psi_l\rangle\otimes|0\rangle_a=\sqrt{p_D}
|u_l\rangle|0\rangle_a+\sqrt{1-p_D}|\phi_l\rangle|1\rangle_a,
\end{equation}
such that the symmetric states $\{|\Psi_0\rangle, |\Psi_1\rangle,
|\Psi_2\rangle, |\Psi_3\rangle\}$ are projected to $\{|u_0\rangle,
|u_1\rangle, |u_2\rangle, |u_3\rangle\}$ with a probability $p_D =
4 \ast |c_{\rm min}|^2$ when a projective measurement on the
ancilla gives the $|0\rangle_a$ state, where $|c_{\rm min}|=
\min\{ |\cos\theta_1|,|\cos\theta_2\sin\theta_1|,
|\cos\theta_3\sin\theta_2\sin\theta_1|, \\
|\sin\theta_3\sin\theta_2\sin\theta_1|\}$. For instance, in case
of angles satisfying $0\leq \theta_1 \leq \pi/3$, $0 \leq \theta_2
\leq 0.3 \pi $ and $0\leq\theta_3 \leq \pi/4$ the minimum
coefficient is $|\sin\theta_3\sin\theta_2\sin\theta_1|$.

In this case, the orthogonal states $|u_l\rangle$ are found to be
the four-dimensional Fourier transform acting on logical states
$|l\rangle$, i.e, these states are given by:
\begin{equation}
|u_l\rangle = {\cal F} \mid l \rangle = \frac{1}{2} \sum_{k=0}^3
e^{i\pi k l/2} |k\rangle,\label{u}
\end{equation}
Hence, the orthogonal states $|u_l\rangle$ are superpositions of
the logical basis. We must apply the inverse of the Fourier
transform for carrying out the discrimination among them in the
logical basis which, in its matrix representation is given by:
\begin{eqnarray}
{\cal F}^{-1}=\frac{1}{2}\begin{pmatrix}
 1&  1& 1 & 1 \\
 1& -i &-1  &i  \\
 1&  -1&  1&  -1\\
 1& i &  -1& -i
\end{pmatrix}. \label{Ucb}
\end{eqnarray}
In terms of linear optics, this transformation can be regarded as
a symmetric eight port beam splitter \cite{Zukowski96}.

\section{Experimental setup with two-photon states}

It is possible to implement the discrimination protocol by using single-photon states,
where the logical states are defined by propagation paths. However, having a controlled
source of single photons is rather difficult. Usually, for this purpose a highly
attenuated pulsed laser is used, with a mean photon number less than one photon per
pulse. For instance, ultralow intensity pulses are used for establishing quantum key
distributions in cryptography experiments \cite{Gisin00}.

Here, we describe an experimental setup for implementing the optimal protocol for
discriminating linearly independent quantum states, by using a simple optical system
based on two-photon states generated in a spontaneous parametric down conversion process.
The optimum is defined in the sense that the protocol maximizes the average success
probability. We codify non-orthogonal quantum states in propagation paths in one of the
down-converted photons (signal) and the other down-converted photon (idler) that will be
used for coincidence measurement, i.e., this photon will ensure the presence of the other
photon in one of the nonorthognal states. Thus, logical state $|j\rangle$, with
$j=0,1,2,3$, corresponds to the $j$-th propagation path of the photon. The discrimination
protocol is divided into four steps: preparation of the symmetric states; conditional
ancilla-system evolution; projective measurement on the ancilla; and finally, in the case
of conclusive measurement, discrimination of an orthogonal system's states.

\begin{figure}[ht]
\includegraphics[width=8.0cm,height=5.0cm]{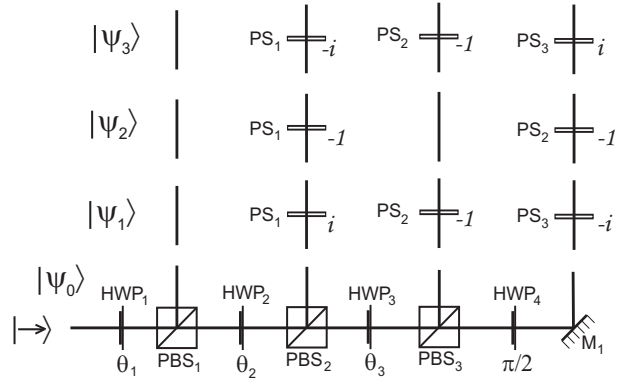}
\caption{Experimental setup for generating symmetric states in Eq. \ref{eq:symstates}. In
all the figures PBS, HWP and PS denote polarized beam splitter, half wave retardation
plates, and phase shifter, respectively. The HWP$_j$ allows for rotating horizontal
polarization in an angle $\theta_j$, with these HWP and PBS the seminal state is
generated, other states are simply generated by inserting PS in the propagation path of
the photon.} \label{fig:states}
\end{figure}
As we have described above, generalized quantum measurements are implemented by embedding
the quantum system into a large Hilbert space by adding an ancilla followed by an
entangling operation. In this protocol, we use the polarization degree of freedom of the
photon as our ancillary system. Hence, in the preparation stage of the symmetric states,
we consider a photon initially prepared with horizontal polarization as input. Using half
wave retardation plates (HWP), polarized beam splitter (PBS) and phase shifters (PS) we
are able to generate the four symmetric states, see Fig.~\ref{fig:states}. The HWP$_i$
rotates the polarization of the photon in an angle $\pi/2-\theta_i$. Hence, the vertical
polarization of the photons is reflected at the PBS$_i$ and this component is used for
defining the $|i-1\rangle$ logical state. The transmitted polarization goes trough the
HWP$_{i+1}$. Actually, it is well known that, by using a HWP, a lossless PBS and a PS
with appropriate parameters, any U(2) transformation can be implemented \cite{Barnett01}.
Considering that we have chosen the values of rotation angles at HWP's such that the
minimum coefficient is $|\sin\theta_3\sin\theta_2\sin\theta_1|$, we generate the
$|\Psi_0\rangle$ state. We remark that HWP$_4$ rotates the polarization of path $4$ from
horizontal to vertical polarization, so that at the end of the preparation stage the
polarization of the photon is factorized from the path states, i.e., in all the
propagation paths the polarization remains vertical. In the same way, other states
$|\Psi_j\rangle$ are generated by inserting phase shifters, see Fig. \ref{fig:states}.

\begin{figure}[ht]
\includegraphics[width=8.0cm,height=5.0cm]{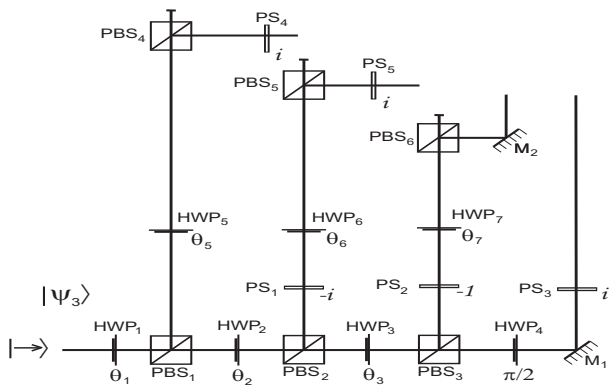}
\caption{Conditional evolution of the ancilla (polarization) depending on the logical
states (propagation path) is achieved by inserting HWP in logical states $|0\rangle$,
$|1\rangle$ and $|2\rangle$. The projection measurement on the ancilla is achieved by
inserting PBS into the same propagation path, so that nonconclusive measurements are
obtained when the photon is transmitted through one of these PBS.}
\label{fig:transformation}
\end{figure}

The first step, in the discrimination protocol, is to apply the conditional evolution
(\ref{U}) onto the symmetric states, which corresponds to a conditional rotation of the
polarization (ancilla) depending on the propagation paths of the photon (logical states).
Hence, the transformation is defined by its action on the logical states and the ancilla
in $|0\rangle$ state:
\begin{eqnarray}
U|00\rangle&=& sin\theta_3 sin\theta_2 tg\theta_1 |00\rangle+\sqrt{1-\left(sin\theta_3sin\theta_2tg\theta_1\right)^2}|01\rangle\nonumber\\
U|10\rangle&=& sin\theta_3  tg\theta_2|10\rangle+\sqrt{1-sin^2\theta_3tg^2\theta_2}|11\rangle\nonumber\\
U|20\rangle&=&  tg\theta_3 |20\rangle+\sqrt{1-tg^2\theta_3}|21\rangle \nonumber \\
U|30\rangle&=&|30\rangle,\label{rotation}
\end{eqnarray}
which is implemented with HWP$_5$, HWP$_6$ and HWP$_7$. The optimum discrimination
process is attained when we choose rotation angles at these HWP as $\theta_5 =
\cos^{-1}(-\tan\theta_1\sin\theta_2\sin\theta_3)$, $\theta_6 =
\cos^{-1}(-\tan\theta_2\sin\theta_3)$ and $\theta_7 = \cos^{-1}(-\tan\theta_3)$. The
projective measurement is implemented right after applying the conditional evolution
given by Eq. (\ref{rotation}). Here, this is done by inserting polarized beam splitters
PBS$_4$, PBS$_5$ and PBS$_6$ in propagation paths $0$, $1$ and $2$, respectively, as is
depicted in Fig.~\ref{fig:transformation}. An inconclusive measurement is obtained when a
photon with horizontal polarization is transmitted through any one of these PBS. If this
projective measurement gives a conclusive measurement (transmission of vertical
polarization) that one of the $\{|u_l\rangle\}$ states has been transmitted, and for
having a full discrimination, we need to determine which one of these $\{|u_l\rangle\}$
states has been transmitted.

Here, all the $\{|u_l\rangle\}$ states are orthogonal superpositions of the propagation
paths, and each one of them is univocally associated with one of the non-orthogonal
states. Hence, the last step in this protocol is the measurement of these orthogonal
states, and for this purpose, it is convenient first to implement a unitary rotation
satisfying $|l\rangle = {\cal F}^{-1}|u_l\rangle$, since in this case the discrimination
is done by a detection of a photon propagating in path $l$. This unitary transformation
is carried out using an eight port interferometer \cite{Zukowski96}.

The above described protocol is easily generalized to the case of $2^M$ symmetric states
to be discriminated. In table \ref{table1} we listed the number of optical component as a
function of the number of non-orthogonal states being discriminated.

\begin{table}
  \centering
\begin{tabular}{|c|c|c|c|c|c|}
\hline Number of states ($2^M$)& HWP & PBS & BS \\
\hline 4 & 7 & 6 & 4 \\
\hline 8 & 15 & 14 & 12 \\
\hline 16& 31 & 30 & 32 \\ \hline
\end{tabular}
\caption{Number of optical components for the discrimination protocol for different
numbers of non-orthogonal states being discriminated. The total number of these
components is approximately given by $2^M(M+2)$. The number of other optical components,
such as mirrors and phase shifters, are of the same order.}\label{table1}
\end{table}

\begin{figure}[ht]
\includegraphics[width=8.0cm,height=5.0cm]{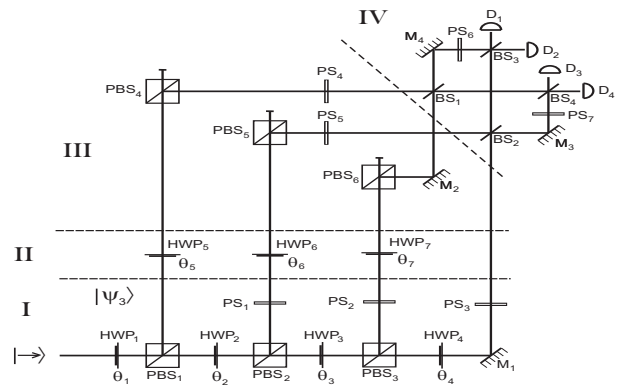}
\caption{General scheme for discrimination of the four symmetric states: (I) Preparation
of state $||Psi_l\rangle$; (II) Conditional evolution of composite system; (III)
Projection measurement; and (IV) detection. Here, the eight port interferometer has been
inserted in the last stage of the experimental setup, which is in the right upper side of
this figure.}\label{esquemageneral}
\end{figure}

We consider using use an Argon ion laser in a continuous wave operation, which pumps a
$\beta$-Barium Borate nonlinear crystal with a power of 350 mW. The laser is made to
operate in a single frequency mode at 351.1 nm, and the presence of other frequencies are
eliminated by inserting a highly dispersive prism right after the laser. In addition, an
interference filter of 10 nm, centered around 351.1 nm, is inserted into the propagation
path of the pump field. Hence, two-photon states with center frequencies at 702.2 nm are
generated. The nonlinear BBO crystal has been cut for SPDC type II, i.e, the propagation
paths of down converted photons are non-collinear. We select signal (idler) photons
linearly polarized in the horizontal (vertical) plane by inserting a Wallaston prism,
with an extinction rate of $100,000:1$.

In the propagation path of the signal photon we insert the setup for implementing the
discrimination protocol. We assume that all the PBS have an extinction rate of 1,000:1.
Controlled rotations of polarization states are accomplished by using HWP. In our case,
the relative angle is adjusted to generate the appropriate coefficients $c_k$ of
symmetric states, Eq. (\ref{eq:symstates}). The purpose of the presence of PS appears to
be evident after the projective measurement, due to the implementation of ${\cal F}^{-1}$
unitary transformations for mapping $|u_l\rangle$ states onto $|l\rangle$ states.

The eight port interferometer must be completely balanced and stabilized, where we deal
with four interferometers in a Mach-Zehnder configuration. This can be done by a phase
adjust mechanism on mirrors $M_1$ to $M_4$ . The angles and positions of these mirrors
must be adjusted to optimize the interference fringes in the four output ports. For this
purpose, mirrors and BS$_1$ to BS$_4$ must be mounted on precision translation stages,
allowing the relative phase between the arms of each the Mach-Zehnder interferometer to
be accurately varied. This stabilization process will be crucial for the discrimination
protocol \cite{Mosley06}. Here, we would like to remark that detectors $D_1$ to $D_4$ in
the signal path are connected with detector $D_i$ in the idler path for coincidence
measurement.

\section{Summary}

We have proposed a scheme for the experimental discrimination of the four symmetric
states. The protocol has been designed for obtaining the optimal value of conclusive
measurements, which is given by the Chefles´s bound. Our scheme considers a reduced
number of optical components and it can easily be generalized to the case of $2^N$
symmetric states. This, to the best of our knowledge, is the first proposal which can be
generalized to larger dimensional quantum systems. The experimental setup is based on
two-photon states from SPDC, which allows us to reach the optimal value for conclusive
discrimination. Hence, by the transmission of a $|\Psi_l\rangle$ state at a time and
coincidence measurement measuring between signal and idler photons, it is possible to
obtain the conclusive probability $p_D$. The main experimental requirement is the
stabilization of interferometers in Mach-Zehnder configurations.

We envisage the employment of the above described setup, for discriminating
non-orthogonal symmetric states, for key distribution in a quantum cryptographic
protocol. Recent works have demonstrated that cryptographic protocols are more robust
against noise channels when using larger dimensional quantum systems \cite{crypto}. For
this purpose, the sender randomly chooses to generate one of the non-orthogonal states.
In this case the propagation paths, after the generation stage, are coupled to single
mode fiber optics, so that the polarization remains constant throughout the fiber. The
receiver implements the discrimination protocol and the cases of conclusive measurement
give a common element of the key to both the sender and the receiver. The presence of an
eavesdropper, in between of authenticated users, can be detected in the authentication
stage, where sender and receiver publicly announce a reduced number of the elements of
the cryptographic key. Alternatively, this presence can also be noticed in a modification
of the probabilities of nonconclusive measurement, which does not require a disclosing of
part of the cryptographic key. This work is under study and we will publish elsewhere the
study on the security of such a protocol. Besides, we also will study applying this
protocol to the problem of discriminating between subsets of non-orthogonal quantum
states, for this problem we will follow the work of Y. Sun {\em et. al} \cite{Sun01},
where the case of a subset from three non-orthogonal states is studied.

\section{ACKNOWLEDGMENT}

We would like to thank Ming Yang for stimulating discussions. This work was supported by
Milenio ICM P02-49F and Grant FONDECYT No. 1061046. X.S. acknowledges financial support
from CONACyT-Mexico and also thanks the hospitality at the Department of Physics,
University of Concepci\'on during her research stay in Concepci\'on.

\end{document}